# Field emission in diode and triode vacuum nanostructures

Michael V. Davidovich[a]

*Saratov State University, P. O. Box 410012, Astrakhanskaya street, 83, Saratov, Russia*

Nikolai A. Buhuev

*JSC "NPP "Almaz", P. O. Box 410012, Panfilova street, 1, Saratov, Russia*

Ravil K. Yafarov

*Saratov branch of IRE RAS, P. O. Box 410019, Zelenaya street, 38, Saratov, Russia*

[a]Electronic mail: davidovichmv@info.sgu.ru

The paper discusses the nano-diode and nano-triode structures of vacuum electronics. Such structures may have the dielectric film with the thickness of several nanometers wich is located on the cathode. Such film with a large dielectric constant reduces the thickness of the potential barrier by about film thickness and reduces the height of barrier. The grid electrode structure may contain several periods of metallization with a thickness of tens of nanometers, which allows one to obtain the resonant under barrier tunneling. For all structures we obtained electrostatic Green's function, the built profiles potential barriers, the calculated tunneling coefficients and Volt-Ampere Characteristics (VAC) with regard to the distribution of electron energies. The structures under consideration require to use of low voltages on the gate (grid) electrodes. They are promising for the electron gun with a large current controlled by low voltage on the grid, for example, in TWT with subsequent acceleration of the electron beam. Since a major influence on the tunneling provides the grid, the electrons emitted from the grid with approximately the Fermi velocity, and the calculation of the electron gun with multiple electrodes after the grid is greatly simplified.

## I. INTRODUCTION

It is well known that to determine the cold emission the Fowler-Nordheim formula

$$J = AE^2 \exp(-B\varphi_0^{3/2}/E) \tag{1}$$

for current density is usually used. Here E is the normal electric field on the cathode and $\varphi_0$ is the working function. This is true for the remote on the infinity of the anode and requires clarification for nanoscale structures. For such structures the longitudinal component of the electric field can depend on the coordinates. If we assume that the main contribution in effect are from the electrons having the Fermi energy, so the current density $J$ is proportional to the coefficient of tunneling $D$. For the simplest triangular approximation of the potential barrier it has the form $D = \exp(-4W^{3/2}\sqrt{2m}/(3\hbar eE))$. To obtain the current density, one must consider the electrons distribution on energy.

It is very important to determine the cold field emission current for the actual configuration of the potential function in planar vacuum diode and triode structures as function of voltage. Great importance is the calculation of emissions in real diode and triode structures. The triode structure with a control electrode allows one to increase significantly the emission current at more small working potentials and reduce the emission threshold. The multielectrode structures with quasiperiodic potential are very interesting for increasing the tunnel current. Such distribution may be produces by multilayered grid structure, in which the resonance tunneling is possible.

In this paper in order to determine the emission characteristics of diode structures we use the multiple image technique relative to the plane of the cathode, dielectric film, grid and anode. The simplest structure with two planes of electrodes is the flat diode[1-3]. There are three planes in diode with dielectric film at the cathode. For triode structures such procedure was made for cathode-grid and grid-anode regions. But for any gap the maximum number of image planes is three, and it equal to two for gap without dielectric film, including the gaps between grid electrodes. In order to avoid singularity we move these reference planes from the boundary of atomic layers for half-period of the lattice *a*. This constant is expressed through the work function. If there is a dielectric or high resistant semiconductor film on the cathode, one must enter additional planes of images[3].

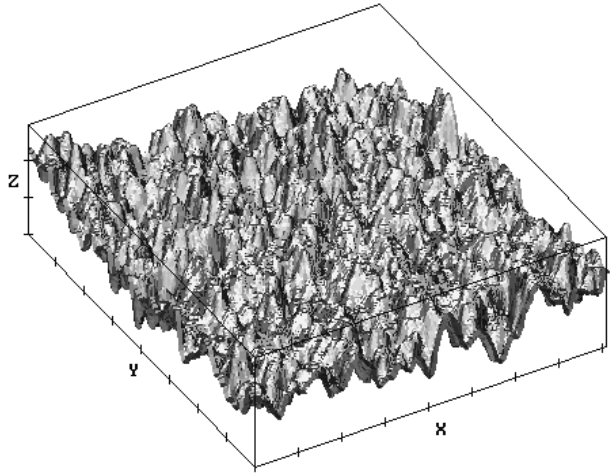

Fig. 1. Atomic force microscope image of nano-diamond film with a thickness of 6 nm on a graphite substrate

The main problem of cold matrix Spindt-type cathode consists in heating the needles. Also such cathode has a small effective working surface. Recent studies show the high effectiveness of the cathodes of the nanocluster diamond-graphite structures with small angles of needles and a large effective surface[4] Fig.1. Such diamond-dielectric phase on graphite substrate or other conductor (for example, glassy-carbon electrode) may be considered as nanoscale dielectric film. Such structures have low work function, so they are very attractive to research.

## II. PROBLEM STATEMENT

To get the quantum mechanical potential function, we use the classic approach and consider the cold field emission in a flat vacuum diode with size d and thin dielectric film with permittivity $\varepsilon$ and thickness $t$. For multielectrode structures the gap potential is constructed similarly to the potential in the diode. The potential energy can be obtained by the method of images[1-3,5], but it turns into infinity at the cathode ($z = 0$) and an anode ($z = d$). The forces of images cease to operate at a distance of the order of the electron shells (half the period of the crystal lattice). So we consider the reduction area $a < z < d-a$ by shifting the reference planes on $a$ (see Fig. 2). Consider, for example secluded cathode. For such solitary case ($d = \infty$) we have the voltage

$$V(z) = \begin{cases} -Ez - e/(16\pi\varepsilon_0(z+a)), & z > 0; \\ V_e = -e/(16\pi\varepsilon_0 a), & z \leq 0. \end{cases} \quad (2)$$

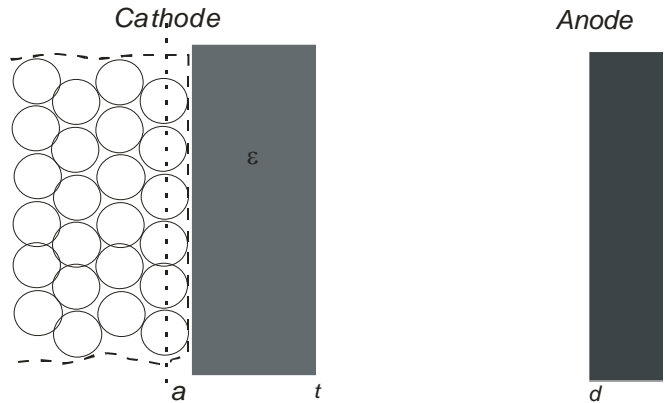

Fig. 2. The schematic configuration of the electrodes and dielectric film in the diode: the reference plane at the cathode is shifted on the distance $a$

From this the potential maximum is $V_m = V(z_0) = Ea - 2\sqrt{eE/(16\pi\varepsilon_0)}$ in the point $z_0 = -a + \sqrt{e/(16\pi\varepsilon_0 E)}$. Also the turning points are

$$z_{1,2} = -(a + U_e/E)/2 \pm \sqrt{(a+V_e/E)^2/4 - e/(16\pi\varepsilon_0 E) - a(V_e/E)}. \tag{3}$$

The voltage is measured from infinity, where the electron is free. To determine the barrier we do not use a quantum-mechanical consideration, and consider, as usual, the electrons in periodic potential in the metal of cathode and anode as free particles. We used the classical approach based on electrostatic Green's function[3,5]. In this case the periodic quantum mechanical potential inside the electrodes (including grid) are constant, and the electrons are free. It is enough to find the shape of two barriers: cathode-grid and grid-anode. The multiple images give the value $W = e^2(16\pi\varepsilon_0)^{-1}(a^{-1} - 2.77/d)$ for barrier height. The electron energy is distributed from zero to the Fermi energy $E_F$. But the anode $U_A$ and grid $U_G$ voltages displace its start of the countdown. When there are quadratic approximation of the potential[1-2], for diode we have: $eV(z) = 4(eU_A - W)(z/d - 1/2)^2 + W + E_F - ezU_A/d$. Here $U_A$ is the anode voltage. It works for small $d$. For arbitrary $d$ one must use the multiple image theory[3,5].

## *A. Multiple images*

The multiple images for the charge at the right of the dielectric layer are presented in Fig. 3. Here the electric flow reflection coefficient[5] is $k = (1-\varepsilon)/(1+\varepsilon)$. The vacuum diode potential function is[1,2]:

$$E_p = eV(z) = -\frac{e^2}{8\pi\varepsilon_0}\left[\frac{1}{2z} + \sum_{n=1}^{\infty}\left(\frac{nd}{(nd)^2 - z^2} - \frac{1}{nd}\right)\right]. \tag{4}$$

As image force does not act at atomic distance ~ $a$, so we must replace $z \to z + a$ in (4) and exclude the infinities (see Fig. 2). The potential energy has the maximum at $z = d/2$, and

$$V(d/2) = -(4\pi\varepsilon_0 d)^{-1} e^2 \sum_{n=1}^{\infty} (-1)^n n^{-1} = -(4\pi\varepsilon_0 d)^{-1} e^2 \ln(2) . \qquad (5)$$

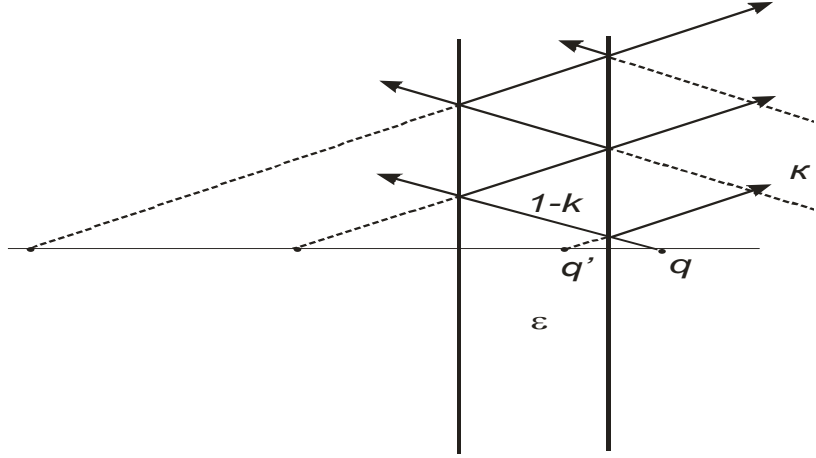

Fig. 3. The schematic configuration of the electric induction flow for charge near the dielectric layer

So, the maximum of work function is

$$W = \frac{e^2}{4\pi\varepsilon_0} \left( \frac{1}{4a} - \frac{0.693}{d} \right) . \qquad (6)$$

There are some approximations of barrier in the flat diode[1,2], for example, the parabolic approximation

$$eV(z) = -\frac{4(W - eV_A)}{d^2}(z - d/2)^2 + W + E_F - \frac{ezV_A}{d},$$

and the approximation for large $d$:

$$eV(z) = -\frac{2^n(W - eV_A)}{d^{2n}}(z - d/2)^{2n} + W + E_F - \frac{ezU_A}{d}.$$

In view of the above, we obtain the exact formula

$$E_p = eV(z) = -\frac{e^2}{16\pi\varepsilon_0}\left[\frac{1}{z+a} + 2\sum_{n=1}^{\infty}\left(\frac{nd}{(nd)^2 - (z+a)^2} - \frac{1}{nd}\right)\right] - \frac{ezU_A}{d}. \qquad (7)$$

Keeping the first members in sum, one has the good approximation:

$$V(z) = -\frac{e}{16\pi\varepsilon_0}\left\{\frac{1}{z+a} + \frac{z}{d-a}\left[\frac{1}{(2d-a)} + \frac{2d(z-2a)}{[d^2 - (z-a)^2](d+a)}\right]\right\} - \frac{zU_A}{d}. \qquad (8)$$

This is a very accurate formula, although the use of accurate formula (7) is not difficult.

## B. Electrostatic Green's function and barrier profile

If we have the Green's function $\Phi(\rho, z, z')$ of complicated structure, the work function may be constructed as $W = e\varphi = -eV_m$, $V(z) = V_m - Ez - \Phi_0(z+a) - \Phi_0(a)$. Here $V_m$ is the maximal potential in such expressions as (8), and $\Phi_0$ is the Green's function without singular term. This term is correspond to charge in the free space and must be removed as the charge does not act on himself. The Green's function construction is based on the free space Green's function $G(\mathbf{r} - \mathbf{r}') = 1/(4\pi R)$. We denote $|\mathbf{r}| = \sqrt{x^2 + y^2 + z^2}$ and $g(\rho, z) = G(x, y, z)$. Then the Green's function of solitary dielectric slab with thickness $2t$ is as follows:

for $z' = t + h$, $z > t$

$$\Phi(\rho, z, z') = \Phi^{++}(\rho, z \mid t, h) = \frac{q}{\varepsilon_0}\left[g(\rho, z - z') + kg(\rho, z - z' + 2h) - k(1 - k^2)\sum_{n=1}^{\infty}k^{2(n-1)}g(\rho, z - z' + 4nt + 2h)\right], \qquad (9)$$

for $z' = t + h$, $z > t$

$$\Phi(\rho, z, z') = \Phi^{++}(\rho, z \mid t, h) = \frac{(1-k^2)q}{\varepsilon_0} \sum_{n=1}^{\infty} k^{2(n-1)} g(\rho, z - z' - 4(n-1)t). \quad (10)$$

For electron the charge $q = e$, so we take $\rho = 0$ and introduce the notation $V(z \mid z')$ for (9) and (10). Our goal is to get the electron potential energy in the plate capacitor with dielectric slab.

### 1. Diode with dielectric slab at the cathode

According (4) the potential of electron located at $z'$ in plate capacitor with dielectric slab without singular term is

$$V(z \mid z') = \Phi_0(0, z, z') - \Phi(0, z, -z') +$$
$$+ \sum_{n=1}^{\infty} [\Phi(0, z, 2nd + z') - \Phi(0, z, -2nd - z') + \Phi(0, z, -2nd + z') - \Phi(0, z, 2nd - z')] \quad (11)$$

Labeling the work required to remove an electron from cathode to the point $z$ as $W(z)$, one can write:

$$W(z) = 0,$$

in the region $0 < z < a'$;

$$W(z) = (e/2)[V(z, z) - V(a', a')],$$

in the region $a < z < t - b$;

$$W(z) = (e/2)[V(t-b, t-b) - V(a, a)],$$

in the region $t - b < z < t + b$;

$$W(z) = (e/2)[V(t-b, t-b) - V(a, a) + V(z, z) - V(t+b, t+b)],$$

in the region $t + b < z < d - a$;

$$W(z) = (e/2)[V(t-b, t-b) - V(a, a) + V(d-a, d-a) - V(t+b, t+b)],$$

in the region $d - a < z < d$. The work is not changing at the distance $a$ near cathode and anode because of absence of image force It is also so at the distance $b$ near the dielectric surface. These lengths are about of atomic distance and are corresponded with work functions of solitary metal and dielectric bodies. We put here for simplicity $b = a$ (the equality of metal and dielectric work functions).

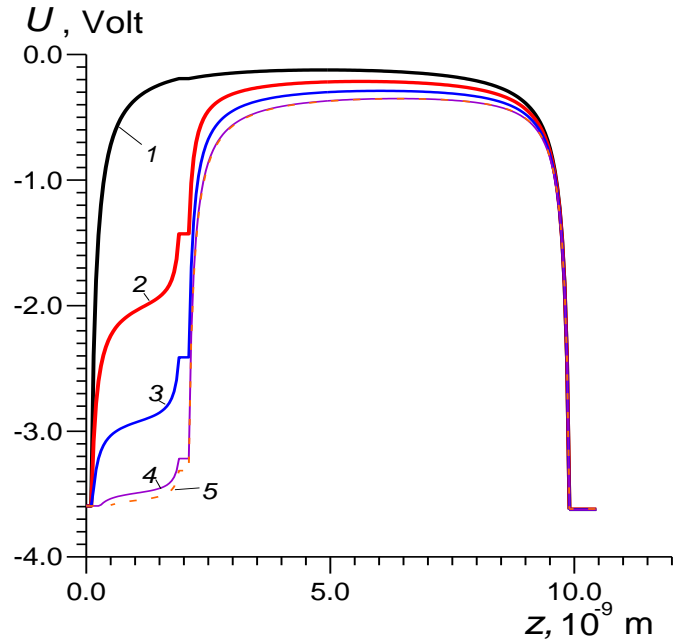

Fig. 4. The potential distribution between the gape of cathode-anode with a dielectric film on the cathode of thickness $t = 2$, for $d = 10$ for different values of the dielectric constant: $\varepsilon = 1$ (curve 1); $\varepsilon = 2$ (2); $\varepsilon = 4$ (3); $\varepsilon = 10$ (4); $\varepsilon = 12$ (5) (dimensions in nm)

## 2. *Multielectrode structure*

In order to calculate potential distribution in the multielectrode structure we use the expressions (7) and (8) for two electrodes without dielectric film and $W(z) = eV_e$ for presence of film. Here $V_e$ is the electrode potential $V_e = U_A$ and $V_e = U_G$ for anode and grid respectively. Some results for potential distribution in diode and triode structures are presented in Fig. 4- 6. The potential for Fig. 6 is counted from Fermi energy level 5 eV at the anode. The results of Fig. 4 and 5 show that the dielectric film reduces the thickness and height of barrier. The influence of the dielectric film on the cathode was noted by Fowler yet in 1931. Such influence is the reason of increasing of tunnel current.

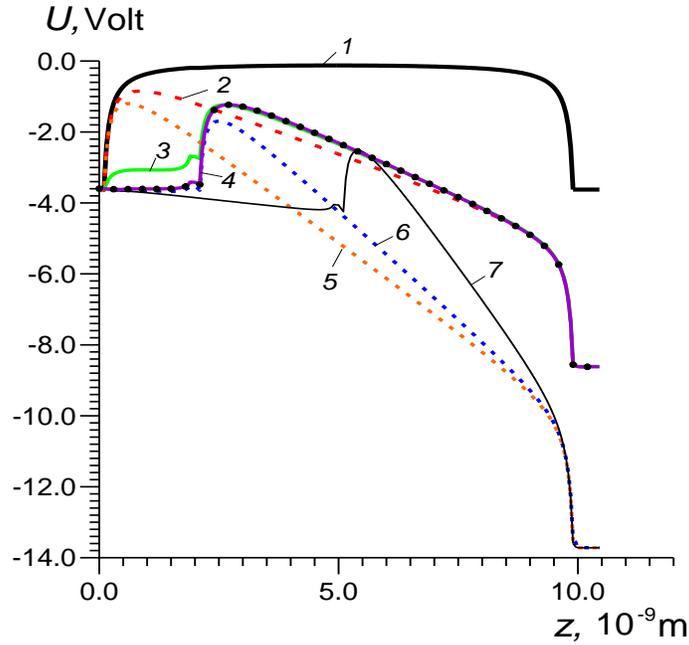

Fig. 5. The potential distribution for different voltages at the anode without film (curves 1, 2, 5), for film with thicknesses $t = 2$ nm (3, 4, 6) and $t = 5$ nm (7): $U_A = 0$ (1); $U_A = 5$ (2, 3, 4); $U_A = 10$ Volt (5, 6, 7); $\varepsilon = 4$ (3); $\varepsilon = 10$ (7); $\varepsilon = 12$ (4, 6)

## III. MODELLING: VOLT- AMPERE CHRACTERISTICS

To determine the current field emission it is necessary: to have the function of the shape of the barrier, to know the tunneling coefficient for this barrier and any kinetic energy $E_{cz}$ of the electron for the normal velocity component, and take into account the electron distribution on energies and the density of states. Although the function of the Fermi-Dirac $f(E)$ allows to obtain the current density at any temperature, we will consider the simplest case $T = 0$.

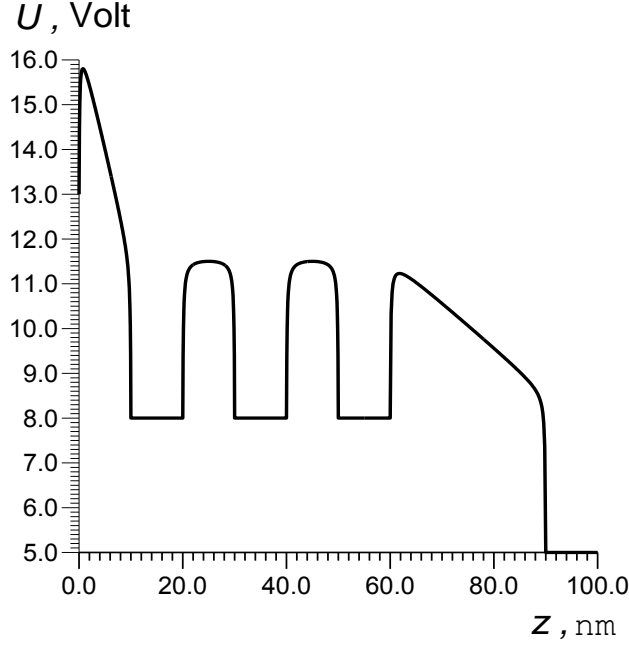

Fig. 6. Calculated profile of barrier for triode with $d = 90$ nm and two-electrode grid with gap and metallization dimensions 10 nm. $U_A = 8$ V, $E_F = 5$ eV

## A. Equations for current density

Denoting $E_{kz}$ and $E_{kr}$ as kinetic energy of the electrons associated with longitudinal and transverse velocities and Fermi-Dirac distribution $f$, we write the density of the tunnel current in the form[1,2]:

$$J = \frac{em}{2\pi^2 \hbar^3} \int_0^{E_{kzm}} D(E_{kz}) dE_{kz} \int_0^{E_{krm}} [f(E) - f(E - eV_A)] dE_{kr} . \qquad (12)$$

We count the potential in Eq. (12) from zero. The total current is the difference between the currents from cathode to anode and the reverse one from the anode to the cathode. So we have for $T=0$:

$$J = \frac{em}{2\pi^2 \hbar^3} \left[ \int_{eV_A}^{E_k^+} (E_k^+ - E_{kz}) D(E_{kz}) dE_{kz} - \int_0^{E_k^-} (E_k^- - E_{kz}) D(E_{kz}) dE_{kz} \right]. \qquad (13)$$

Here $D(E_{kz})$ is the coefficient of tunneling, $E_k^- = E_F$, $E_k^+ = E_F + eU_A$ and $E_0^+ = eU_A$ are the maximal and minimal kinetic energy at the cathode. Potential is quasiperiodic if the grid has several periodically located electrodes as for potential Fig. 6, which is calculated for three periods.

## B. Current approximation

To obtain current density (12), one must numerously solve the Schrodinger equation for a complicated potential barrier. To avoid this we have approximated the tunneling coefficient in the form $D(E_{kz}) = 1 - A(1 + \beta E_{kz})\exp(-\alpha E_{kz})$. It is expedient to introduce different factors: $\alpha_+$ - for current from the cathode, and $\alpha_-$ - for current from the anode. Note that we calculated these coefficients for the real given shape of the potential barrier and $U_A$. The proposed approximation allows us to analytically calculate the integral (13), but the result is quite cumbersome. Here is a simpler result for the case $\beta = 0$. Then there are three tunneling coefficients: $D(eU_A + E_F) = 1 - A\exp(-\alpha_+(eU_A + E_F))$, $D(eU_A) = 1 - A\exp(-\alpha_+ eU_A)$, $D(E_F) = 1 - A\exp(-\alpha_- E_F)$, which must be calculated numerically. Then we have such definitions of $\alpha_\pm$, $\tilde{\alpha}_+$:

$$\alpha_+ = \ln\left(\frac{1 - D(0)}{1 - D(E_F + eU_A)}\right)/(E_F + eU_A) = (E_F + eU_A)^{-1}\eta_0,$$

$$\tilde{\alpha}_+ = (eU_A)^{-1}\ln\left(\frac{1 - D(0)}{1 - D(eU_A)}\right) = (eU_A)^{-1}\eta_1,$$

$$\alpha_- = E_F^{-1}\ln\left(\frac{1 - D(0)}{1 - D(E_F)}\right) = E_F^{-1}\eta_2.$$

Dimensionless positive parameters $\eta_0$, $\eta_1$, $\eta_2$ determine in this case the shape of the barrier. For direct current it is advisable to use a weighted average value: $\alpha = [\alpha_+(E_F + eU_A) + \tilde{\alpha}_+ eU_A]/(E_F + 2eU_A)$. If $eU_A \sim E_F$ or more, it is possible to neglect the reverse current. In this case

$$J = \frac{em}{2\pi^2\hbar^3}\left[(E_F + eU_A)\int_{eU_A}^{E_F + eU_A} D(E_{kz})dE_{kz} - \int_{eU_A}^{E_F + eU_A} E_{kz}D(E_{kz})dE_{kz}\right].$$

We introduce the following dimensionless functions:

$$F(\alpha, W) = 1 - \exp(-\alpha W)$$

$$f(\alpha, W_1, W_2) = (\alpha W_2)^{-1} \exp(-\alpha W_1) F(\alpha, W_2),$$

$$g(\alpha, W_1, W_2) = (\alpha W_2)^{-2} \exp(-\alpha W_1)[(\alpha W_1 + 1) F(\alpha, W_2) - \alpha W_2 \exp(-\alpha W_2)].$$

Then the calculation of the integral (13) with their use determines the VAC in the form

$$J(V_A) = A \frac{emE_F^2}{2\pi^2 \hbar^3} \left[ g(\alpha, eV_A, E_F) - g(\alpha_-, 0, E_F) - \left(1 + \frac{eV_A}{E_F}\right) f(\alpha, eV_A, E_F) + f(\alpha_-, 0, E_F) \right]. \quad (14)$$

For low anode voltage $eV_A \ll E_F$, because $eV_A \ll E_F$ and $A \approx 1$, we have the result

$$J(V_A) = \frac{emE_F^2}{2\pi^2 \hbar^3} \left[ \left(1 + \frac{eV_A}{E_F}\right)^2 \Psi(\alpha(E_F + eV_A)) - \frac{(eV_A)^2}{E_F^2} \Psi(\alpha eV_A) - \Psi(\alpha_- E_F) \right]. \quad (15)$$

Here is the function $\Psi(x) = x/6 - x^2/24 + x^3/120$.

## IV. RESULTS AND DISCUSSION

The results of calculations of volt-ampere characteristics for diode and diode structures are presented in Fig. 7. All curves have been calculated using the Eq. (14). For testing we used the direct calculation of the integral with ten energy values (points of integration) for barrier in diode. The results are in good accordance. You can specify a large value of achievable current densities. Of course, there are some limitations: the need to consider the actual temperature and heat, space charge, nonstationarity tunneling and other factors. For complicated barriers the above approximation is insufficient, and it is necessary to calculate the integrals taking into account the solutions of the Schrödinger equation for each point of integration.

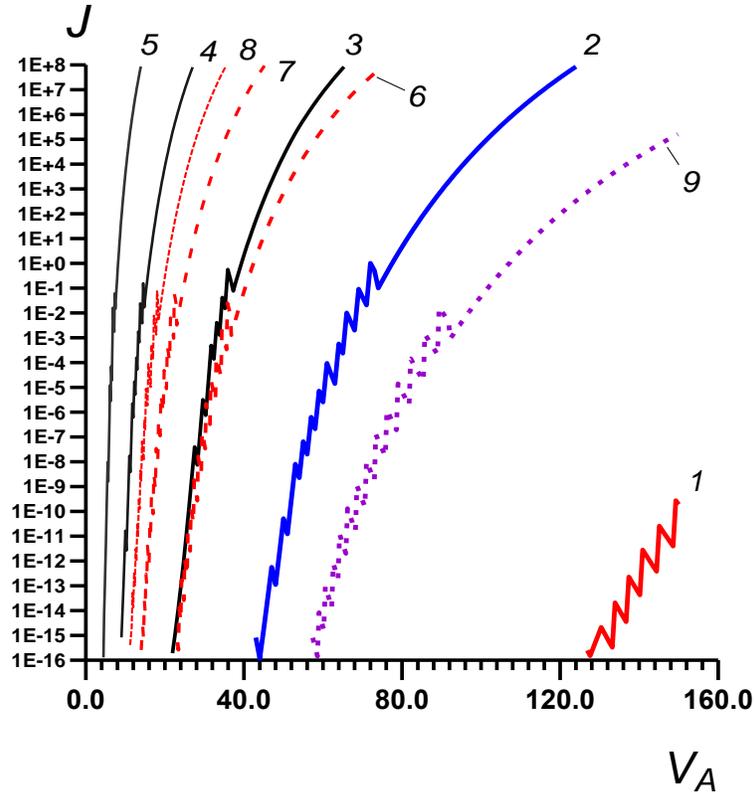

Fig. 7. The density of the tunneling current (A/m$^2$) depending on the voltage at the anode (V) for diode structures with different sizes of the gap (curves 1, 2, 3, 4, 5) and triode structures with size $d$=100 nm $U_G/U_A = 0.5$ (curves 6, 9); 0.8 (7); 1.0 (8)

## A. Experiments

The experimental measurements of the current density for diamond-graphite structure of Fig. 1 are presented in Fig. 8. In this figure tunnel current increases with the film thickness of the nano-diamond phase. This film has a dielectric constant in the range of 10.

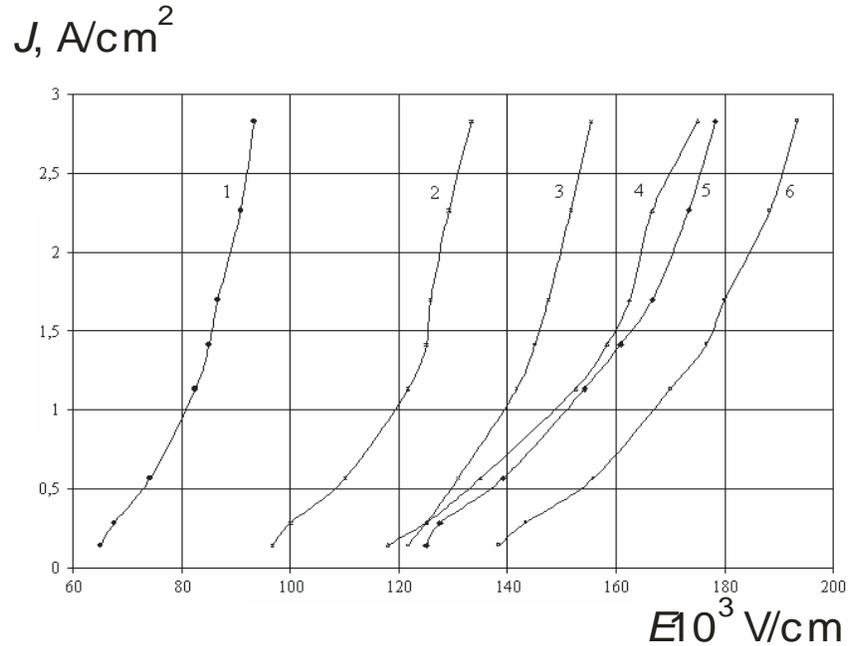

Fig. 8. The experimental current density in the diode with nano-diamond film on graphite substrate: $t = 16, 12, 10, 8, 4, 2$ nm (curves 1, 2, 3, 4, 5, 6 respectively)

## *B. Discussion*

Experimental results confirm the theoretical conclusion about the impact of dielectric film on emissions. The experiment was performed for diode structures with the distance of the cathode - anode an order of magnitude greater than the calculation, which limits the ability to compare results. However, the growth trend current with increase in film thickness occurs. In this case, it varied from 2 to 12 nm. Recently for the thickness of 80 nm we have obtained a current density much more. Obviously, the current density should reach saturation with increase film thickness. For such film it is better to use semiconductor high-resistance material.

## V. SUMMARY AND CONCLUSIONS

The expressions for determining of the shape of the potential barrier and for calculation the tunneling current for the structure of the cathode-anode (diode) and the cathode-grid-anode (triode), including those containing dielectric film on the cathode have been obtained. It is shown that such film can significantly reduce the work function and to reduce the width of the barrier, which leads to an increase of the tunneling current. The effect manifests itself most strongly when the film thickness is an essential part of the barrier. In this respect it differs significantly from the size effect in nanoscale structures (quantum wells), for which work function decreases with decreasing of sizes.

## ACKNOWLEDGMENTS

The paper was presented at the Tenth International Vacuum Electronic Sources Conference ( IVESC-2014), St. Petersburg, 29.06 - 04.07, 2014. The work was supported by the Ministry of education and science of the Russian Federation in the framework of the project of the state tasks in the field of scientific activity No 3.1155.2014/K.